\documentclass[a4paper,conference]{IEEEtran}
\IEEEoverridecommandlockouts

\usepackage{geometry}
\usepackage{amsmath,amssymb,amsfonts}
\usepackage{mathptmx}
\usepackage{amssymb}
\usepackage{algpseudocode}
\usepackage[linesnumbered,ruled]{algorithm2e}
\usepackage{array}

\newtheorem{lem}{Lemma}[section]

\usepackage{array}
\usepackage{stfloats}
\usepackage{blindtext}
\usepackage{subcaption}
\usepackage{textcomp}
\usepackage{stfloats}
\usepackage{url}
\usepackage{verbatim}
\usepackage{graphicx}
\usepackage{cite}
\usepackage{graphicx}
\usepackage{textcomp}
\usepackage{xcolor}
\def\BibTeX{{\rm B\kern-.05em{\sc i\kern-.025em b}\kern-.08em
    T\kern-.1667em\lower.7ex\hbox{E}\kern-.125emX}}
\begin{document}
\title{Experimental Demonstration of Remote Synchronization in Coupled Nonlinear Oscillator
}
\author{\IEEEauthorblockN{Sanjeev Kumar Pandey}
\IEEEauthorblockA{\textit{Electrical Engineering Department}
\textit{IIT Delhi, Hauz Khas}\\
New Delhi, India (110016) \\
sanjeev@ee.iitd.ac.in}
}
\maketitle
\begin{abstract}
 This study investigates remote synchronization in scale-free networks of coupled nonlinear oscillators inspired by synchronization observed in the brain's cortical regions and power grid. We employ the Master Stability Function (MSF) approach to analyze network stability across various oscillator models. Synchronization results are obtained for a star network using linearization techniques and extended to arbitrary networks with benchmark oscillators, verifying consistent behavior. Stable synchronous solutions emerge as the Floquet multiplier decreases and the MSF becomes negative. Additionally, we demonstrate remote synchronization in a star network, where peripheral oscillators communicate exclusively through a central hub, drawing parallels to neuronal synchronization in the brain. Experimental validation is achieved through an electronic circuit testbed, supported by nonlinear ODE modeling and LTspice simulation. Future work will extend the investigation to arbitrary network topologies, further elucidating synchronization dynamics in complex systems.
\end{abstract}

\section{Introduction}
Synchronization is a ubiquitous phenomenon observed in various natural and artificial systems. Pioneering work by Winfree explored synchronization in large populations of coupled oscillators \cite{winfree1967biological}. Since then, the field has grown significantly, encompassing models from Kuramoto \cite{10313029} and Winfree motivated by biological processes \cite{marvel2009invariant}, such as brain waves \cite{frank2000towards}, cardiac pacemaker cells \cite{peskin1975mathematical}, firefly flashing \cite{buck1968mechanism}, and biochemical oscillations  \cite{strogatz1993splay}.

Remote synchronization is a more recent discovery \cite{olmi2024multilayer,luo2024effects,cui2024exponential,wei2024enhancing}. Bergner et al. investigated phase synchronization in star networks of non-identical Stuart-Landau oscillators \cite{bergner2012remote}, establishing necessary conditions and showing that fixed-amplitude systems cannot exhibit this behavior. Minati et al. further explored remote synchronization and pattern formation in larger rings of nonlinear electronic oscillators \cite{minati2015remote}. The growing interest in brain abnormalities has made remote synchronization in the brain a topic of significant interest. Motivated by this, Qin et al. investigated remote synchronization and pattern formation in star networks using the Kuramoto-Sakaguchi model \cite{qin2018stability}, studying the role of structural connections between neurons in synchronizing cortical regions. All these study have been done for star network and bipartite networks but do not comment on the arbitrary network.

This investigation aims to achieve remote synchronization in arbitrary networks of coupled nonlinear oscillators. To ensure a comprehensive exploration of this phenomenon, a multifaceted approach encompassing analytical, numerical, and experimental methods has been adopted. For the analytical component, the Floquet theory, the Master Stability Function approach, and the Gershgorin circle theorem have been employed to investigate the underlying mechanisms and conditions for remote synchronization. 

\section{Mathematical Background}
\subsection{Floquet Theory}
The study of synchronization in Linear Time-Varying (LTV) systems is of considerable interest across various scientific disciplines due to the ubiquity of time-varying parameters in real-world scenarios. Floquet theory, which enables the analysis of periodic systems, offers a vital analytical framework for investigating the stability and synchronization characteristics of LTV systems \cite{pandey2024synchronization}. 

\begin{lem}\label{Flo}
 Consider a linear time-varying system governed by the equation $\dot {x}=A(t)x$, where $A(t+T)=A(t)$ for all $t$ and $T$ is a positive constant. Let $\phi(t,t_{0})$ denote the state transition matrix that maps the state at time $t_{0}$ to the state at time $t$, let $\phi(t+T, t)$ be the associated Floquet matrix. The Floquet matrix exhibits the following properties:
 \begin{itemize}
     \item Existence and Periodicity: The Floquet matrix is periodic with period $T$.
     \item Eigenvalues and Eigenvectors: The eigenvalues of the Floquet matrix, known as Floquet multipliers, and their corresponding eigenvectors form a robust and complete set of solutions to the time-varying system.
     \item Stability Criterion: The stability of the time-varying system is determined by the absolute values (moduli) of the Floquet multipliers.
 \end{itemize}
\end{lem}

\subsection{Master Stability Function}
The Master Stability Function (MSF) is a mathematical tool used to analyze synchronization in networks of coupled dynamical systems (Algorithm 1). It systematically determines the stability of the synchronized state across various coupling strengths. The MSF approach linearizes the system's dynamics around the synchronized state and calculates the largest Lyapunov exponent. A negative MSF value indicates stability, while a positive value suggests instability \cite{aristides2024master,pandey2024synchronization}.

\begin{algorithm}
\caption{Master Stability Function Approach}
\KwIn{
    $f(\mathbf{x})$: System dynamics function \\
    $H(\mathbf{x})$: Coupling function \\
    $G$: Laplacian coupling matrix \\
    $K$: Coupling strength
}
\KwOut{Plot of $\text{MSF}(\mu_{max})$ vs. $\gamma$ for stability analysis}

\tcc{Network Setup}
$\dot{\mathbf{x}}_i = \mathbf{F}(\mathbf{x}_i) + K \sum_{j=1}^N \mathbf{G}_{ij} \mathbf{H}(\mathbf{x}_j)$ \tcp*[r]{Coupled dynamics (Eq. 2)}
$\dot{\mathbf{x}}_i = \mathbf{F}(\mathbf{x}_i)$ \tcp*[r]{Individual dynamics ($\mathbf{x}_i \in \mathbb{R}^d$)}
\tcp{Zero-sum property for undirected graphs:} $\sum_{j=1}^n a_{ij} = 0$ for all $i$

\tcc{Linearization}
$\mathbf{x}_s(t)$ \\
$\mathbf{J} = \left. \frac{\partial \mathbf{F}}{\partial \mathbf{x}} \right\rvert_{\mathbf{x}_s(t)}$ \tcp*[r]{Jacobian of $\mathbf{F}$ at $\mathbf{x}_s$}
$\mathbf{DH}_s = \left. \frac{\partial \mathbf{H}}{\partial \mathbf{x}} \right\rvert_{\mathbf{x}_s(t)}$ \tcp*[r]{Jacobian of $\mathbf{H}$ at $\mathbf{x}_s$}

\tcc{Variational Equation}
$\dot{\delta \mathbf{x}}_i = \mathbf{J} \delta \mathbf{x}_i + K \sum_{j=1}^n \mathbf{G}_{ij} \mathbf{DH}_s \delta \mathbf{x}_j $  (for each $i$) \tcp*[r]{Linearized dynamics (Eq. 3)}

\tcc{Block Diagonalization and Dimensionality Reduction}
$\mathbf{Q}, \mathbf{\Lambda} \gets$ Eigendecomposition of $\mathbf{G}$ \tcp*[r]{$\mathbf{\Lambda}$: diagonal matrix of eigenvalues $\lambda_i$}
$\delta \mathbf{y} = \mathbf{Q}^{-1} \delta \mathbf{x}$ (Transform to block-diagonal coordinates)

\tcc{Calculate MSF from Blocks}
\For{each eigenvalue $\lambda_i$ in $\mathbf{\Lambda}$ (except $\lambda_1 = 0$)}{
  \For{$K$ in $[K_{\min}, K_{\max}]$}{
    $\gamma \gets -K \lambda_i$ \;
    $\mathbf{M} \gets \mathbf{J} + \gamma \mathbf{DH}_s$ \;
    $\mu_{\max} \gets$ Largest Lyapunov exponent of $\dot{\zeta} = \mathbf{M} \zeta$ \;
    $\text{MSF}[\gamma] \gets \mu_{\max}$ \;
  }
}
\end{algorithm}

\section{Remote synchronization results}
\begin{figure}[ht]
\centering
\includegraphics[width=0.33\textwidth]{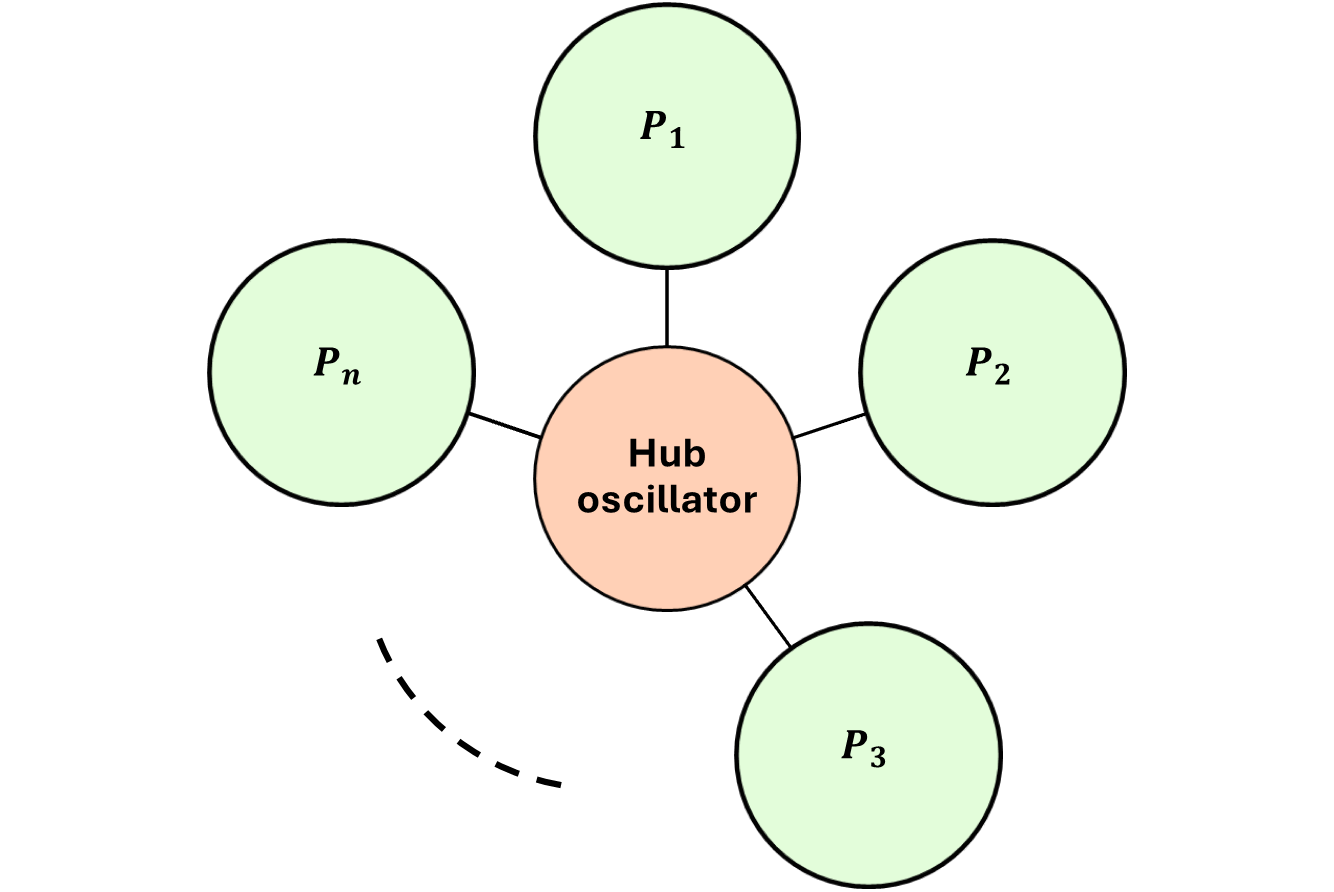}
\caption{Star topology with central Hub and Peripheral node P.}
\label{fig:9}
\end{figure}

Synchronization between neuronal populations, facilitated by the thalamus, is essential for information transmission across cortical regions. This synchronization, arising from the collective oscillations of numerous neurons, enhances functional connectivity. The underlying neuronal structure resembles a network, with the thalamus acting as a relay to synchronize disparate cortical areas. Figure \ref{fig:9} illustrates this concept, depicting the central oscillator as the thalamus and peripheral oscillators as neurons from various cortical regions, forming a scale-free network.
Scale-free networks are defined by their structural invariance undergrowth, characterized by a power-law degree distribution: $P_{d e g}(\mathbf{k}) \propto k^{-\gamma}$. This scale-invariant property confers robustness, clustering, phase transitions, fractal dimension, and self-organized criticality to these networks.\\
\subsubsection{Example First: Kuramoto oscillator}
The dynamics of coupled nonlinear oscillators are
\begin{equation} \label{remote_1}
\begin{split}
\begin{array}{l}
\dot{\theta}_{0}=\omega_{0}+\sum_{i=1}^{n} \varphi_{i} \sin \left(\theta_{i}-\theta_{0}\right), i \in \mathcal{N}_{n} \\
\dot{\theta}_{i}=\omega+A_{i} \sin \left(\theta_{0}-\theta_{i}\right) \quad 
\end{array}
\end{split}
\end{equation}
  In this model, $\theta_{0}$  denotes the phase of an oscillator, $\omega_{0}$ is the natural frequency of the central oscillator, $\omega$  is the natural frequency of the peripheral oscillators, $\varphi_{i}$ signifies the incoming coupling strength, and $A_{i}$ represents the outgoing coupling strength of the coupled oscillators \cite{qin2020mediated}. Remote synchronization manifests in a star network when a subset of peripheral oscillators achieves phase synchronization while the phase of the central mediator, responsible for their interconnection, remains unsynchronized.
  
Considering $n+1$ oscillators coupled by star network, where $n \geq 2.$ Where
$\theta_{i}$ is the phase of oscillator,
$\omega_{0}$ is the natural frequency of the hub oscillator,
$\omega$ is the natural frequency of the peripheral oscillator,
$\varphi_{i}$ is the incoming coupling strength, and 
$A_{i}$ is the outgoing coupling strength.
Defining new variable $x$ 
$$
x=\theta_{0}-\theta_{i}
$$
On taking the derivative of $x$ and putting the values from the equation of dynamics of oscillators, it will
$$
\begin{array}{c}
\dot{x}_{i}=\omega_{0}+\sum_{i=1}^{n} \varphi_{i} \sin \left(\theta_{i}-\theta_{0}\right)-\left[\omega+A_{i} \sin \left(\theta_{0}-\theta_{i}\right)\right] \\
\dot{x_{i}}=\omega_{0}-\omega-\sum_{i=1}^{n} \varphi_{i} \sin \left(\theta_{0}-\theta_{i}\right)-A_{i} \sin \left(\theta_{0}-\theta_{i}\right) \\
\dot{x_{i}}=\omega_{0}-\omega-\sum_{i=1}^{n} \varphi_{i} \sin \left(x_{i}\right)-A_{i} \sin \left(x_{i}\right)
\end{array}
$$
Here $n=3$ and $i=1,\ldots, n$, so the previous equation then $\dot{x}_{i}$ will be
\begin{equation*}
    \begin{aligned}
\dot{x_{1}}=\omega_{0}-\omega-(\varphi_{1}+A_{1}) \sin (x_{1})-\varphi_{2} \sin(x_{2})-\varphi_{3} \sin (x_{3}) \\
\dot{x_{2}}=\omega_{0}-\omega-(\varphi_{2}+A_{2}) \sin (x_{2})-\varphi_{1} \sin (x_{1})-\varphi_{3} \sin(x_{3}) \\
\dot{x_{3}}=\omega_{0}-\omega-(\varphi_{3}+A_{3}) \sin (x_{3})-\varphi_{2} \sin(x_{1})-\varphi_{3} \sin (x_{2})
\end{aligned}
\end{equation*}
Calculated equilibrium values are $x_{1}^{*}=2 \pi, x_{2}^{*}=2 \pi, \& x_{3}^{*}=2 \pi$, by assuming the system is identical $\left(\mathrm{\varphi}_{1}=\mathrm{A}_{1}=\right.$ $1, \mathrm{~\varphi}_{2}=\mathrm{A}_{2}=1, \mathrm{~\varphi}_{3}=\mathrm{A}_{3}=1$ ). The obtained jacobean matrix is given by
\begin{equation*}
\hspace*{-0.1cm}
\begin{aligned}
{\left[\begin{array}{l}
\delta \dot{x}_{1} \\
\delta \dot{x}_{2} \\
\delta \dot{x}_{3}
\end{array}\right]=\left[\begin{array}{ccc}
-2 \cos x_{1} & -\cos x_{2} & -\cos x_{3} \\
-\cos x_{1} & -2 \cos x_{2} & -\cos x_{3} \\
-\cos x_{1} & -\cos x_{2} & -2 \cos x_{3}
\end{array}\right]\left[\begin{array}{l}
\delta x_{1} \\
\delta x_{2} \\
\delta x_{3}
\end{array}\right]} \\
{=\left[\begin{array}{lll}
-2 \cos (2 \pi) & -\cos (2 \pi) & -\cos (2 \pi) \\
-\cos (2 \pi) & -2 \cos (2 \pi) & -\cos (2 \pi) \\
-\cos (2 \pi) & -\cos (2 \pi) & -2 \cos (2 \pi)
\end{array}\right]\left[\begin{array}{l}
\delta x_{1} \\
\delta x_{2} \\
\delta x_{3}
\end{array}\right]} \\ 
    \end{aligned}
\end{equation*}

\begin{equation} \label{remote_4}
{\left[\begin{array}{l}
\delta \dot{x}_{1} \\
\delta \dot{x}_{2} \\
\delta \dot{x}_{3}
\end{array}\right]=\left[\begin{array}{lll}
-2 & -1 & -1 \\
-1 & -2 & -1 \\
-1 & -1 & -2
\end{array}\right]\left[\begin{array}{l}
\delta x_{1} \\
\delta x_{2} \\
\delta x_{3}
\end{array}\right]}
\end{equation}
The calculated eigenvalues, $-4$, $-1$, and $-1$, all possess negative real parts, indicating system stability. However, if the equilibrium values were odd numbers, the eigenvalues would become positive, leading to system instability.

\begin{figure}[htp]
\centering
\includegraphics[width=0.4\textwidth]{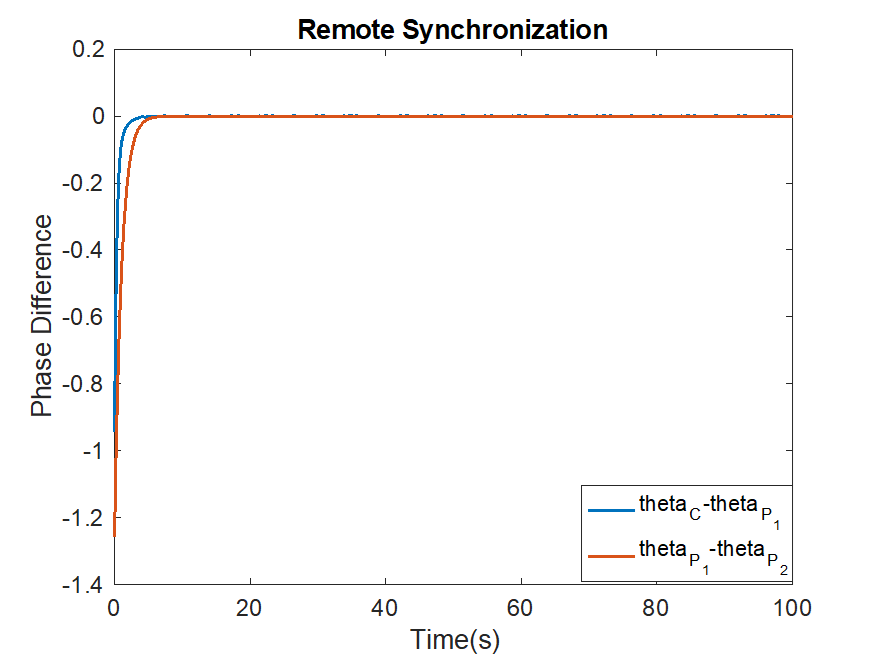} 
\caption{Numerical simulations demonstrate the convergence of phase differences to zero between the central and peripheral oscillators and among the peripheral oscillators themselves.}
\label{fig:10}
\end{figure}

\textbf{ODE simulation of remotely synchronized oscillators}\\
The simulation results demonstrate that all peripheral oscillators achieve synchronization in the presence of the central oscillator. Fig. \ref{fig:10} illustrates the synchronization of frequency and phase among all peripheral oscillators, with the central oscillator exhibiting similar frequency and phase characteristics. Fig. \ref{fig:10} further reveals that the phase difference between peripheral oscillators is zero, and their trajectories converge to zero. The phase difference between the central (hub) and peripheral oscillators also converges to zero, indicating complete phase synchronization.

The Jacobian matrix of \ref{remote_4}, being invertible and diagonally dominant, allows the application of the Gershgorin circle theorem to investigate eigenvalue bounds and system stability \cite{long2024low}. This theorem reveals that all eigenvalues of the Jacobian matrix inherently have positive real parts, implying their location in the right half-plane. However, a negative sign shifts these eigenvalues to the left half-plane, ensuring invertibility. With these foundational concepts established, we now proceed to present our results.
$$ J\left(x^{*}\right)=-\left.\frac{\partial f}{\partial x}\right|_{x=x^{*}}$$
The column diagonal dominance of the Jacobian matrix implies that all eigenvalues of $J\left(x^{*}\right)$ have negative real portions, proving the local asymptotic stability of the equilibrium $x_{i}^{*}$. 
\begin{figure}[ht]
\centering
\includegraphics[width=0.4\textwidth]{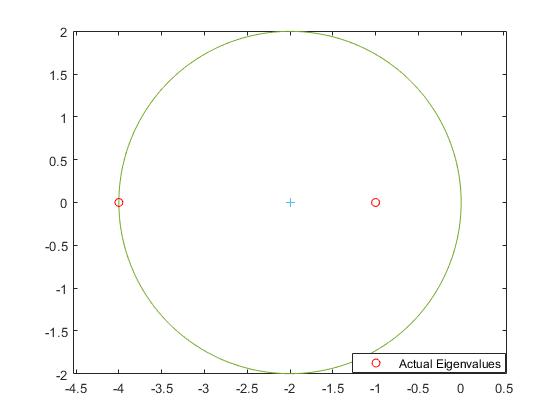}
\caption{Figure illustrates the Gershgorin discs corresponding to the Jacobian matrix.}
\label{fig:11}
\end{figure}

Subsequently, MATLAB was employed to conduct an ODE simulation to calculate bounds. The results indicate that the union of both bounds, defined across all three eigenvalues ($\lambda_{1}=-4$, $\lambda_{2}=-1$, $\lambda_{3}=-1$), lies entirely within the left half-plane, thereby confirming their boundedness. The Greshgorin theorem further corroborates this observation. Consequently, it can be concluded that the system under consideration exhibits both boundedness and stability for this specific configuration.\\
\subsubsection{Example Second: Coupled FHN oscillators using}
The dynamics of coupled oscillators are
\begin{equation} \label{remote_fhn_2}
\begin{split}
\begin{array}{l}
\frac{d v_{H}}{d t}=c\left[v_{H}-w_{H}-\frac{1}{3} v_{H}^{3}\right]+\mathrm{I}+\varphi_{1}\left(v_{1}-v_{H}\right)+\varphi_{2}\left(v_{2}-v_{H}\right) \\
\frac{dw_{H}}{dt}=d\left(a+bv_{H}-w_{H}\right)+\varphi_{1}\left(w_{1}-w_{H}\right)+\varphi_{2}\left(w_{2}-w_{H}\right) \\
\frac{d v_{1}}{dt}=c\left[v_{1}-w_{1}-\frac{1}{3} v_{1}^{3}\right]+\mathrm{I}+A_{1}\left(v_{H}-v_{1}\right) \\
\frac{dw_{1}}{dt}=d\left(a+b v_{1}-w_{1}\right)+A_{1}\left(w_{H}-w_{1}\right) \\
\frac{dv_{2}}{d t}=c\left[v_{2}-w_{2}-\frac{1}{3} v_{2}^{3}\right]+\mathrm{I}+A_{2}\left(v_{H}-v_{2}\right) \\
\frac{d w_{2}}{dt}=d \left(a+bv_{2}-w_{2}\right)+A_{2}\left(w_{H}-w_{2}\right)
\end{array}
\end{split}
\end{equation}
The model assigns the variable $v$ to represent the membrane potential and the variable $w$ to account for the biological neuron's channel dynamics and the neural dynamics of the biological neuron. The parameter values employed in the simulation are: $a = 3$, $b = 2$, $c = 1$, $d = 0.2799991$, $I = 2.1$. The coupling strength $\varphi$ is assumed to be identical in both directions, i.e., from the central oscillator to the peripheral oscillators and vice-versa. Under a star network configuration \cite{cui2024exponential}, one central oscillator is connected to two peripheral oscillators.

\begin{figure}[ht]
\centering
\includegraphics[width=0.47\textwidth]{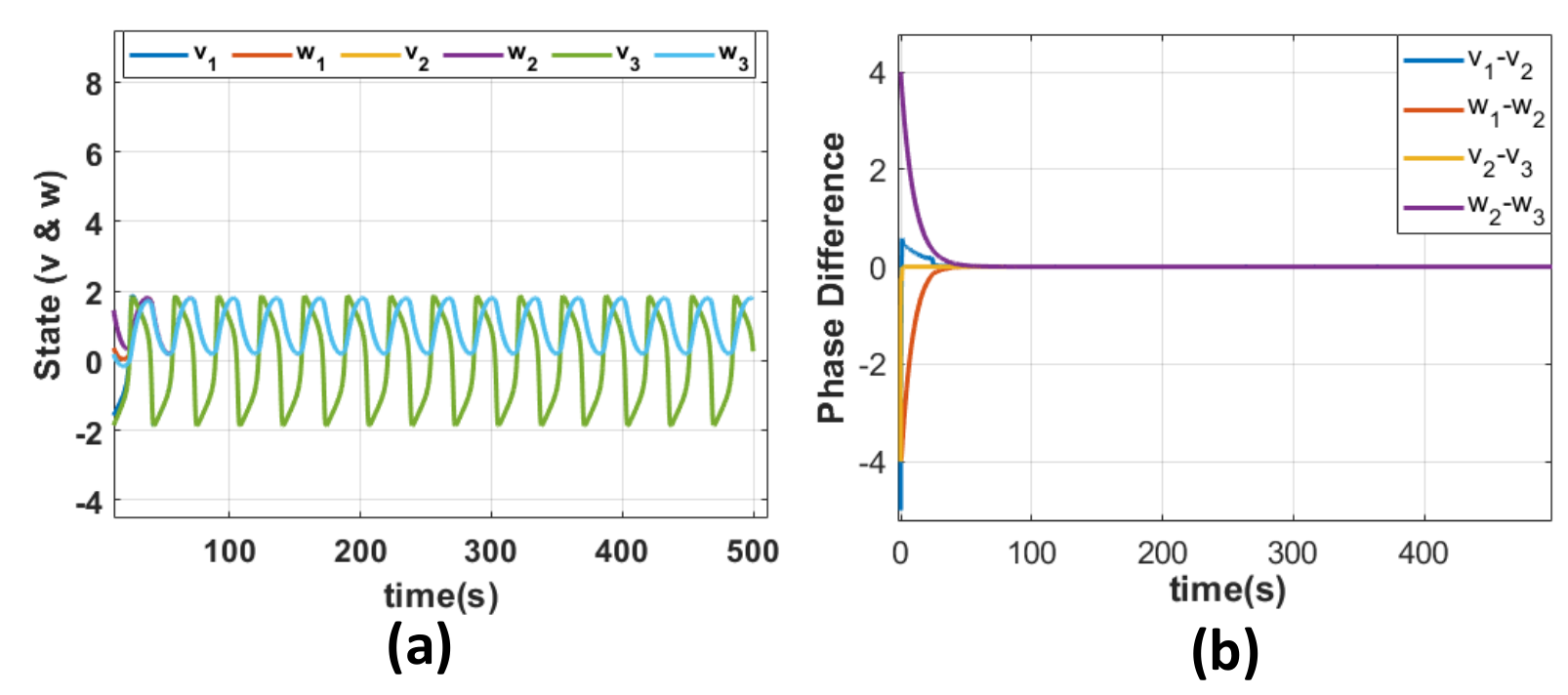}
\caption{Fig. (a) illustrates complete synchronization between the mediator (hub oscillator) and mediated oscillators (peripheral oscillators). Fig. (b) depicts the convergence of phase differences between oscillators to zero, indicating phase synchronization.}
\label{fig:12}
\end{figure}

\textbf{ODE simulation of remotely synchronized oscillators}\\
The simulation results in this section illustrate the synchronization of all peripheral oscillators in the presence of a central oscillator. Fig. \ref{fig:12}(a) illustrates the synchronization of frequency and phase among all peripheral oscillators, with the central oscillator exhibiting similar characteristics. Fig. \ref{fig:12}(b) further reveals zero phase difference between peripheral oscillators, their trajectories converging to zero, and the phase difference between the central (hub) and peripheral oscillators converging to zero.

Equation \ref{remote_fhn_2} has been expanded to explicitly represent the hub and peripheral oscillators, with A denoting the coupling strength from the hub to the peripheral oscillators and $\varphi$ representing the coupling strength from the peripheral to the hub oscillators. The following parameter values were assigned: $a = 3$, $b = 2$, $c = 1$, $d = 0.2799991$, $I = 2.1$, and $A_{1}=A_{2}=\varphi_{1}=\varphi_{2}=0.115$. Subsequently, the system was linearized about its limit cycle, resulting in the Monodromy matrix (M) and given by 

\small\begin{equation} \label{remote_fhn_3}
\begin{aligned}
    =\left[\begin{array}{cccccc}
\zeta_{H} & -c & \varphi_{1} & 0 & \varphi_{2} & 0 \\
db & -d-\varphi_{1}-\varphi_{2} & 0 & \varphi_{1} & 0 & \varphi_{2} \\
A_{1} & 0 & \zeta_{1} & -c & 0 & 0 \\
0 & A_{1} & db & -d-A_{1} & 0 & 0 \\
A_{2} & 0 & 0 & 0 & \zeta_{2} & -c \\
0 & A_{2} & 0 & 0 & d b & -d-A_{2}
\end{array}\right]
\end{aligned}
\end{equation}

In this context, $\zeta_{H}= c(1-v_{H}{ }^{2})+\varphi_{1}+\varphi_{2}$, $\zeta_{1}= c(1-v_{1}{ }^{2})-A_{1}$, and $\zeta_{2}= c(1-v_{2}{ }^{2})-A_{2}$. The Floquet multipliers were computed numerically following the procedures outlined in the mathematical tools section. The resulting Floquet multipliers are$\mu_{1}=1.0 + 0.000i$, $\mu_{2}=0.0 + 0.092i$, $\mu_{3}=0.0 - 0.092i$, $\mu_{4}=0$, $\mu_{5}=0$, and $\mu_{6}=0$. Given that one multiplier equals one and the remaining multipliers have magnitudes less than one, we can deduce that the system exhibits stability.
 \section{Remote synchronization of arbitrary network}
The FHN oscillator has been selected as the nonlinear oscillator to investigate the phenomenon of remote synchronization within an arbitrary network configuration. The dynamical behavior exhibited by an isolated, uncoupled FHN oscillator is described as follows
\begin{equation} \label{eq1}
\begin{split}
\begin{array}{c}
\dot v_{1} =c(v_{1}-w_{1}-\frac{1}{3} v_{1}^{3})+I \\
\dot w_{1}=d(a+bv_{1}-w_{1})\\
\end{array}
\end{split}
\end{equation}
The structural configuration of the arbitrary network under consideration is illustrated below
\begin{figure}[ht]
\centering
\includegraphics[width=0.3\textwidth]{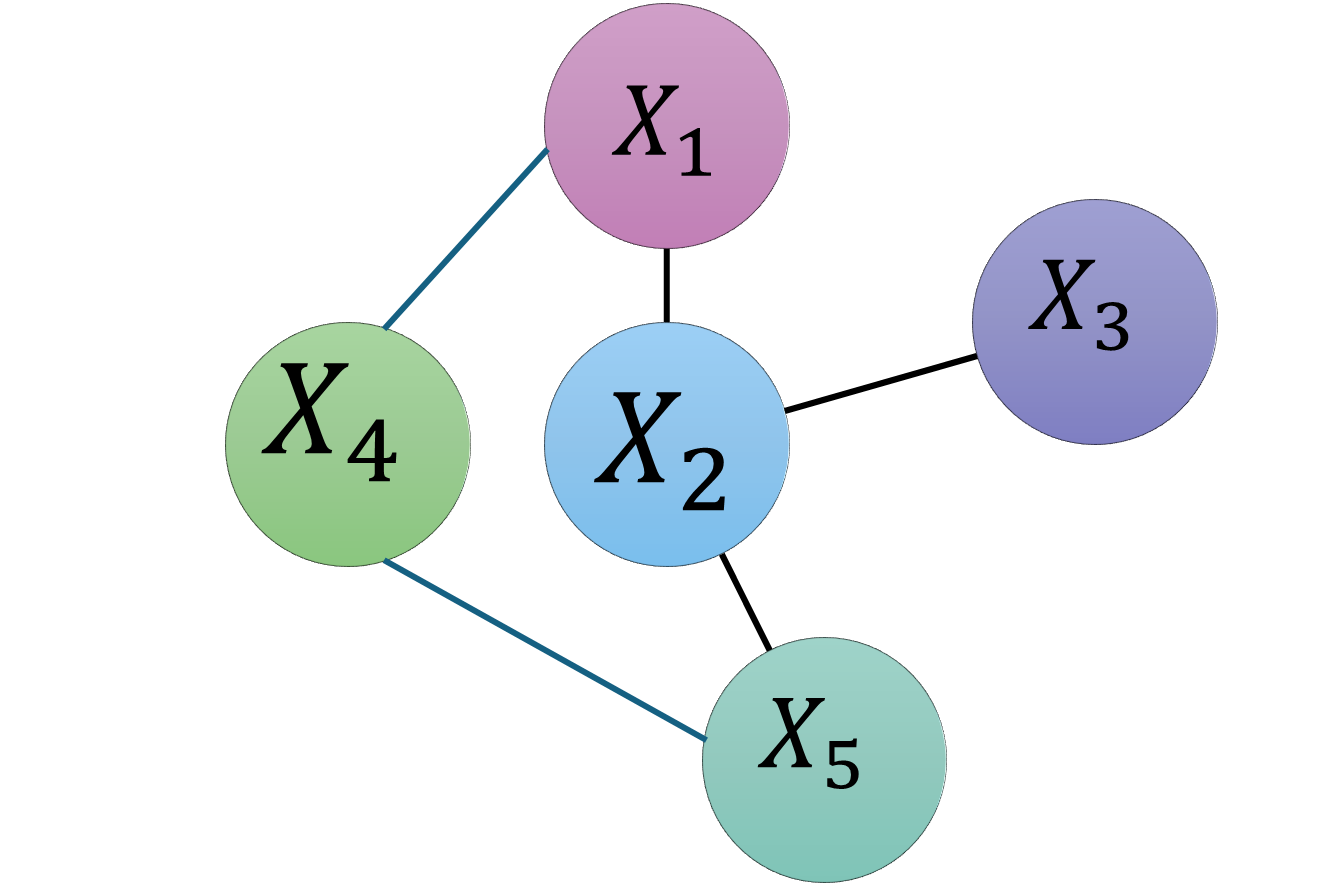}
\caption{Graph with five nodes}
\label{fig:graph}
\end{figure}

\begin{figure}[ht]
\centering
\includegraphics[width=0.48\textwidth]{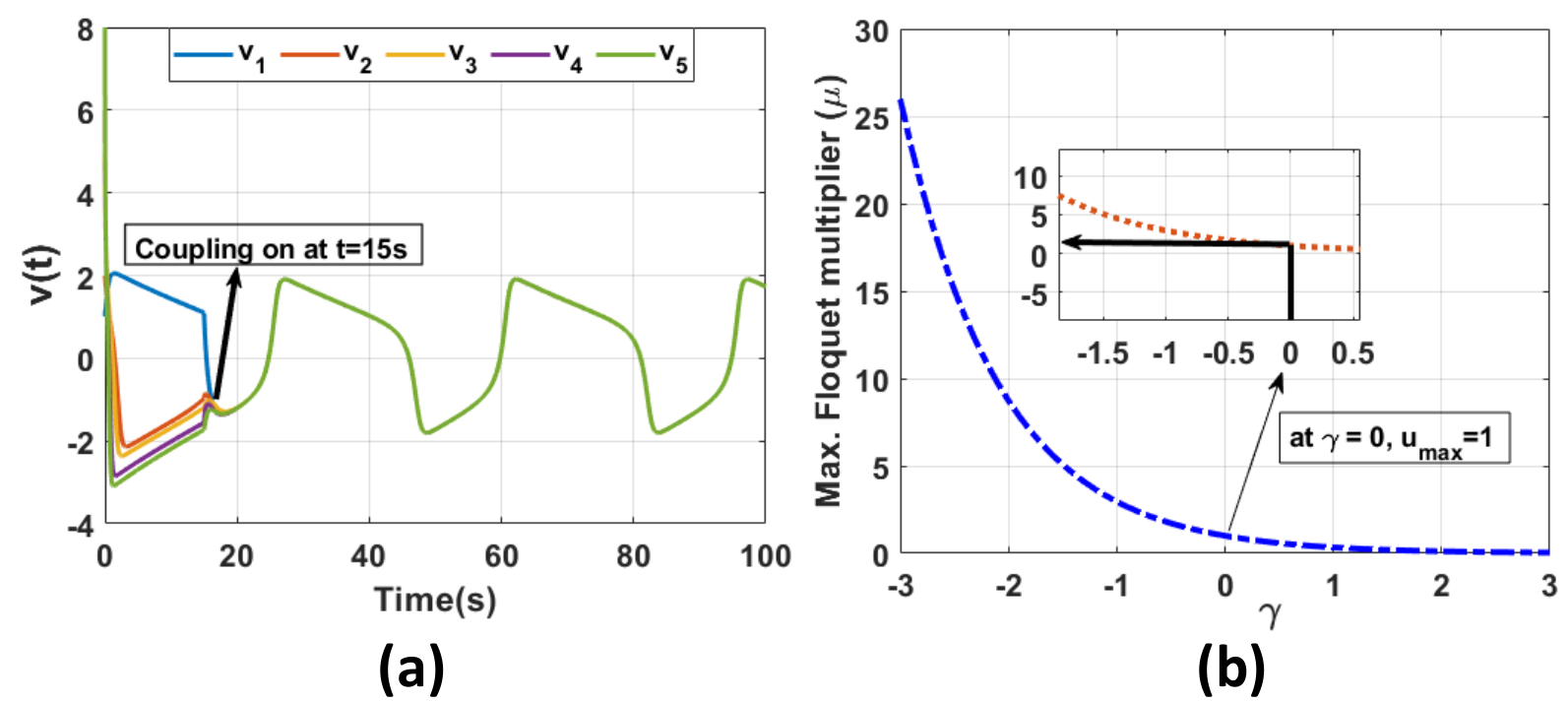}
\caption{Synchronization dynamics within an arbitrary network of coupled FHN oscillators. (a) Illustrates the occurrence of remote synchronization among five oscillators. (b) Depicts the relationship between the Floquet multiplier and the product of coupling strength $\varphi$ and eigenvalues of the diagonalized matrix representation of graph G.}
\label{fig:arb_clus}
\end{figure}

The graph reveals the absence of direct connections between specific node pairs: $X_{3}$ and ($X_{1}$, $X_{5}$,$X_{4}$); $X_{1}$ and $X_{5}$; $X_{4}$ and $X_{2}$. However, indirect communication is facilitated through intermediary nodes, enabling system-wide synchronization. Figure (\ref{fig:arb_clus}) illustrates this, and the coupled oscillator dynamics can be mathematically expressed as follows.

\begin{equation} \label{coupled_eq}
\begin{split}
\begin{array}{l}
\dot v_{1} =c(v_{1}-w_{1}-\frac{1}{3} v_{1}^{3})+I+\varphi(v_{2}+v_{4}-2v_{1}) \\
\dot w_{1}=d(a+bv_{1}-w_{1})+\varphi(w_{2}+w_{4}-2w_{1}) \\
\dot v_{2} =c(v_{2}-w_{2}-\frac{1}{3} v_{2}^{3})+I+\varphi(v_{5}+v_{3}+v_{1}-3v_{2}) \\
\dot w_{2}=d(a+bv_{2}-w_{2})+\varphi(w_{5}+w_{3}+w_{1}-3w_{2}) \\
\dot v_{3} =c(v_{3}-w_{3}-\frac{1}{3} v_{3}^{3})+I+\varphi(v_{2}-v_{3}) \\
\dot w_{3}=d(a+bv_{3}-w_{3})+\varphi(w_{2}-w_{3}) \\
\dot v_{4} =c(v_{4}-w_{4}-\frac{1}{3} v_{4}^{3})+I +\varphi(v_{5}+v_{1}-2v_{4})\\
\dot w_{4}=d(a+bv_{4}-w_{4}+\varphi(w_{5}+w_{1}-2w_{4})) \\
\dot v_{5} =c(v_{5}-w_{5}-\frac{1}{3} v_{5}^{3})+I+\varphi(v_{4}+v_{2}-2v_{5}) \\
\dot w_{5}=d(a+bv_{5}-w_{5}+\varphi(w_{4}+w_{2}-2w_{5})) \\
\end{array}
\end{split}
\end{equation}

The associated Laplacian matrix is defined as follows
\begin{equation} \label{eq3}
\begin{split}
\mathrm{L}=\left[\begin{array}{ccccc}
2 & -1 & 0 & -1 & 0 \\
-1 & 3 & -1 & 0 & -1 \\
0 & -1 & 1 & 0 & 0  \\
-1 & 0 & 0 & 2 & -1 \\
0 & -1 & 0 & -1 & 2
\end{array}\right]
\end{split}
\end{equation}
The eigenvalues resulting from the computation on the Laplacian matrix are as follows: $$ \lambda_{1}= 0.0, \lambda_{2}=0.8, \lambda_{3}=2.0, \lambda_{4}=2.6 , \text { and } \lambda_{5}=4.4$$ and eigenvectors are
$$
\left[\begin{array}{ccc}
\begin{array}{rrrrr}
0.4472 & -0.2560 & 0.7071 & 0.2422 & -0.4193 \\
0.4472 & 0.1380 & 0.0000 & 0.5362 & 0.7024 \\
0.4472 & 0.8115 & -0.0000 & -0.3175 & -0.2018 \\
0.4472 & -0.4375 & 0.0000 & -0.7031 & 0.3380 \\
0.4472 & -0.2560 & -0.7071 & 0.2422 & -0.4193
\end{array}
\end{array}\right].$$

\begin{equation} \label{eq4}
\begin{aligned}
&\dot{y}=\left[\begin{array}{cc}
c\left(1-v_{1}^{2}\right) & -1 \\
d b & -d
\end{array}\right]- \varphi\lambda\left[\begin{array}{cc}
1 & 0 \\
0 & 1
\end{array}\right] {y} \\
\end{aligned}
\end{equation}

Employing the maximum eigenvalue of the Laplacian graph ($\lambda=4.4$) in conjunction with equation \ref{eq4}, the maximum Floquet multiplier is determined to be $u_{1}=1.00$, with the remaining multiplier being $u_{2}=0.24$ for a positive coupling strength. This observation concludes that the system demonstrates stability for this specific coupling strength value across the maximum eigenvalue of the diagonalized Laplacian matrix. The stability of the synchronous state is further supported by the negative value of the master stability function at $\varphi \lambda_{max}$. This is visually corroborated in the plot depicting the maximum Floquet multiplier concerning $\varphi \lambda_{i}$, where $i$ ranges from $-4$ to $4$. Figure \ref{fig:arb_clus} illustrates the decrease in the Floquet multiplier of the master stability equation as the product of coupling strength and eigenvalues of the diagonalized matrix increases. Notably, for negative values of $\varphi \lambda_{i}$, the Floquet multiplier exceeds one, indicating system instability. However, for positive values of $\varphi \lambda_{i}$, the master stability function is negative, resulting in a stable system.\\

\section{Experimental results}
This section details experimental results on remote synchronization in a star topology of coupled nonlinear oscillators. A phase oscillator was initially designed and simulated using LTspice. Subsequently, a physical circuit was constructed with three peripheral oscillators linearly coupled to a central one, enabling communication between each peripheral oscillator and the central one but not directly among themselves. The circuit design exploits the nonlinear gain-voltage relationship in a Wien-bridge oscillator with diode-based amplitude stabilization.

The feedback loop, employing an operational amplifier and a Wien-bridge RC network, determines the oscillation frequency $(f=\frac{1}{2 \pi R C})$. Nonlinearity is introduced via anti-parallel diodes, modeled by the Shockley equation, rendering the amplifier gain $(G)$ dependent on the output voltage $(v_o)$: $G(v_o) \approx G_0-a v_{0}^2$. Oscillations commence when $G>3$, but diode action reduces gain with increasing amplitude, stabilizing at $G(v_o) \approx 3$, forming a negative feedback loop. Notably, temperature significantly impacts diode saturation current $(I_{s})$, influencing oscillation amplitude \cite{sze2021physics}. The feedback current is
$i_{F} \approx \frac{V_{D}}{R_{f_{2}}}+I_{s}\left(e^{\frac{v_{D}}{n V_{T}}}-e^{\frac{-v_{D}}{n V_{T}}}\right)$ and its Taylor series approximation is $i_{F} \approx \frac{v_{D}}{R_{f_{2}}}+2I_{s}\left[\frac{v_{D}}{nV_{T}}+\left(\frac{v_{D}}{3!n V_{T}}\right)^3+\ldots\right]$. Assume that $R_{s}=0 \Omega$, the output voltage of the ideal op-amp is $v_{o} \approx \left(\frac{R_{f_2}}{R_{f_1}}+1\right) v-\frac{I_S R_{f_2}}{3\left(n V_T\right)^3}\left(G_0-1\right)^3 v^{3}$.  
For steady-state with $G \approx 3$,the amplifier gain is $G \approx G_0-\frac{8 I_s R_{f_2}}{9\left(n V_T\right)^3} v_o^2$ and the complete mathematical model is
\begin{equation}
    \frac{d^{2}v_{o}}{dt^{2}}+\omega_{0}\left[\left(3-G_{0}\right)+\frac{8I_{S} R_{f_{2}}}{9\left(n V_{T}\right)^3} v_{o}^{2}\right] \frac{dv_{o}}{dt}+\omega_{0}^{2} v_{o}=0.
\end{equation}
\begin{center}
 TABLE 1: Numerical ratings of used electronic components
  \begin{tabular}{ | l | l | l | p{1.5cm} |}
    \hline
    Symbol & Component  & Value & Units  \\ \hline 
    $R $ & resistor & $1.0\pm5\%$ & $k \Omega$ \\ \hline
    $C $ & capacitor & $1.0\pm10\%$ & $ \mu F$ \\ \hline
    Op-Amp  & UA741CN  &  &  \\ \hline
    $V$ & voltage source  & $\pm 12$ &v  \\ \hline
    \end{tabular}
     \label{table:1}
 \end{center}

\begin{figure}[ht]
\centering
\includegraphics[width=0.45\textwidth]{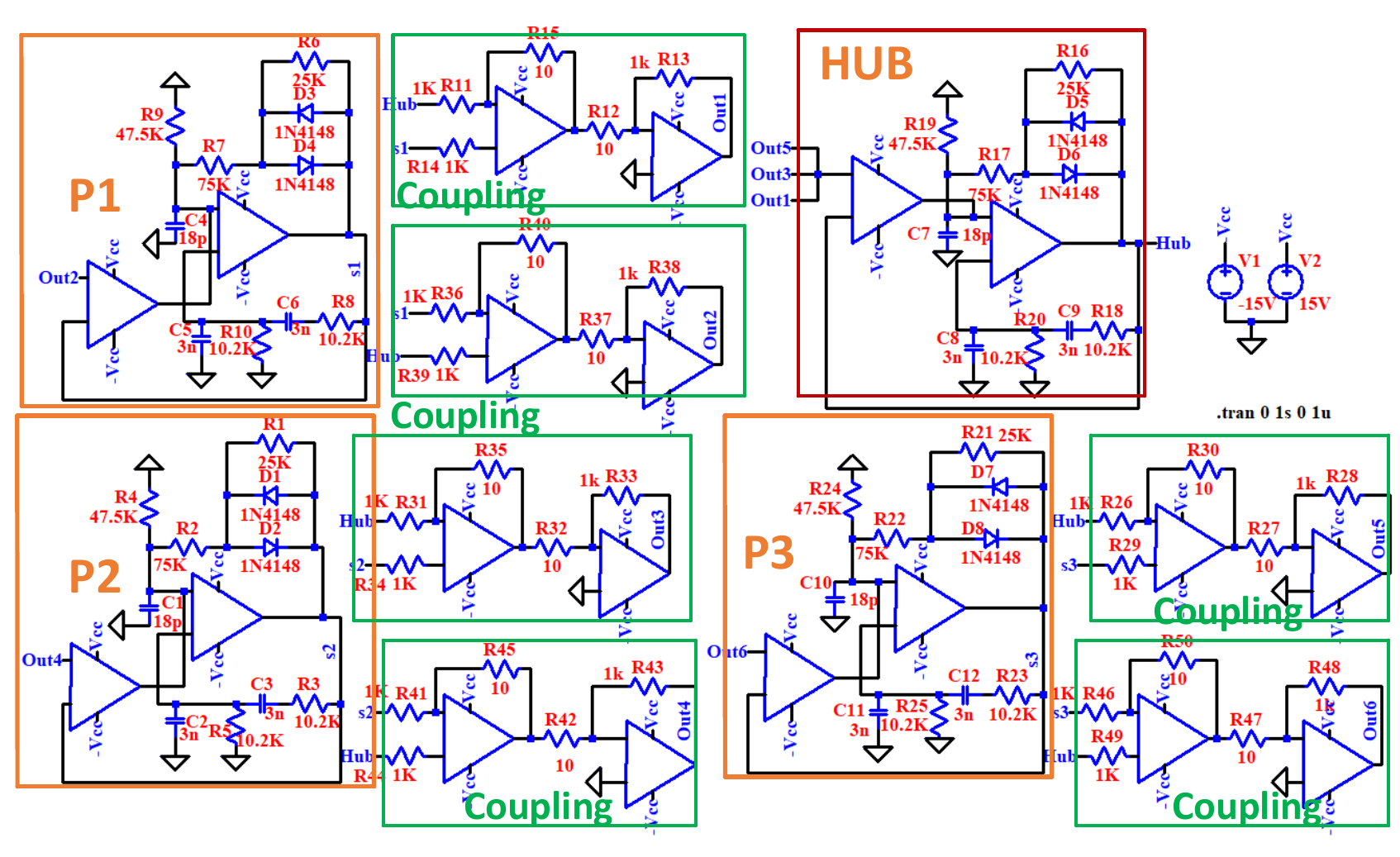}
\caption{Circuit diagram of the coupled phase oscillator.}
\label{fig:17}
\end{figure}

A network comprising three oscillators was employed to demonstrate the phenomenon of remote synchronization. The circuit diagram for the coupled oscillators adheres to the configuration depicted in Figure \ref{fig:17}, with the specific components utilized being enumerated in Table 1. Three peripheral oscillators were linearly coupled to a central oscillator in an all-to-all topology, wherein each peripheral oscillator communicated with the central oscillator but not directly with each other. In Figure \ref{fig:17}, $P1$,$P2$, and $P3$ represent the peripheral oscillators, while "Hub" denotes the central oscillator. The``coupling terms" in the diagram indicate the interconnections between the oscillators.

\begin{figure}[ht]
\centering
\includegraphics[width=0.35\textwidth]{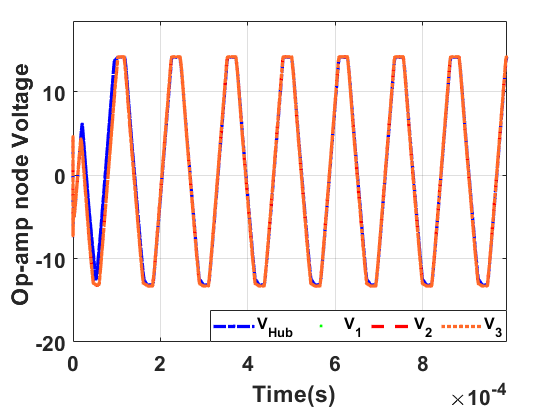}
\caption{Numerical simulation outcomes for a system of coupled nonlinear phase oscillators.}
\label{Star_Coupled_LTSPICE_Simulation}
\end{figure}

\begin{figure}[ht]
\centering
\includegraphics[width=0.33\textwidth]{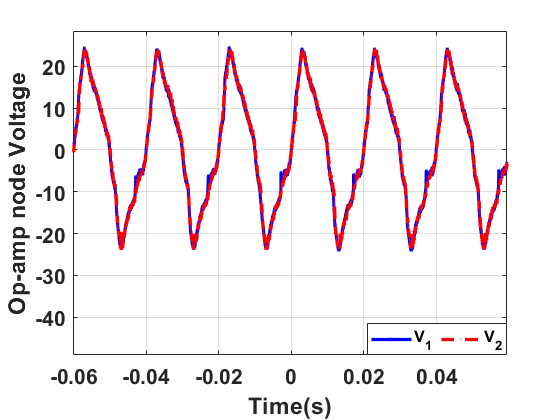}
\caption{Experimental validation of remote synchronization achieved utilizing the components specified in Table 1. The data was acquired via an oscilloscope and subsequently plotted using MATLAB. In this context, $v_{1}$ signifies the node voltage of the central (hub) oscillator, while $v_{2}$ represents the node voltage of a peripheral oscillator.}
\label{fig:18}
\end{figure}

The numerical simulation of coupled oscillators is shown in Fig. \ref{Star_Coupled_LTSPICE_Simulation} using LTspice simulation. Where all the peripheral oscillators get synchronized after the coupling gain is activated.
Electronic components were used to design the experimental setup, as mentioned in Table 1. The biasing voltage in the op-amp was $\pm 12V$, the non-inverting pin was grounded, and the inverting pin was connected to the output of the op-amp of the previous node, thus forming the overall feedback. The component's details are mentioned in Table 1. Experimental behavior has been studied and shown in Fig. \ref{fig:18}. It is observed that all the peripheral and Hub oscillators get synchronized with a single phase and frequency, resulting in synchronization. 

\section{Conclusion}
In conclusion, this study has successfully investigated remote synchronization in scale-free networks of coupled nonlinear oscillators inspired by the synchronization observed in the brain's cortical regions. By employing the Master Stability Function approach, we have analyzed the stability of these networks across various oscillator models. Our results demonstrate that a negative Master Stability Function and decreasing Floquet multipliers indicate stable synchronous solutions. We have also provided experimental validation of remote synchronization in a star network using an electronic circuit testbed, highlighting the potential relevance of our findings to neuronal synchronization in the brain. Future work will extend this investigation to arbitrary network topologies, further deepening our understanding of synchronization dynamics in complex systems.
\section*{Acknowledgments}
The author is grateful to Prof. S. Sen and Prof. I. N. Kar of the control and automation group for their unwavering support and guidance. I am also grateful to S. Biradar and R. M. Bora for their valuable time and suggestions.
\bibliographystyle{ieeetr}
 \bibliography{myreferences}
\end{document}